\documentclass{WileyMSP-template}
\usepackage{comment}
\usepackage{xcolor}
\usepackage{cite}
\usepackage{amsmath} 
\usepackage{ragged2e}
\usepackage{caption}
\usepackage{footnote}
\usepackage[colorlinks=true,linkcolor=blue,citecolor=blue,urlcolor=blue]{hyperref}
\makesavenoteenv{figure}
\begin{document}

%\pagestyle{fancy}
%\rhead{\includegraphics[width=2.5cm]{vch-logo.png}}

\title{Additive Manufacturing-Facilitated Blow Molding for Functional Thin-Walled Polymeric Structures}

\maketitle

% Author: Please give full first and last names for authors and include * after the name of all corresponding authors

\author{Junyu Chen}
\author{Dotan Ilssar*}
\author{Dennis M. Kochmann*}

% Dedication

% Affiliations: Please provide adacemic titles (Prof. or Dr.) for all authors where applicable, and include an institutional email address for all corresponding authors
\begin{affiliations}
Junyu Chen, Dr. Dotan Ilssar, Prof. Dennis M. Kochmann\\
Mechanics \& Materials Laboratory, Department of Mechanical and Process Engineering, ETH Zurich\\
Email Address: dilssar@ethz.ch, dmk@ethz.ch
\end{affiliations}

% Keywords: Please provide a minimum of three and a maximum of seven keywords, separated by commas


% Abstract should be written in the present tense and impersonal style (i.e., avoid we), and be at most 200 words long
\begin{abstract}
\justify
Thin-walled structures capable of large, reversible deformation are key to multistable structures, origami, kirigami, and soft robotics. However, conventional fabrication techniques---including 3D printing, casting, and laser cutting---suffer from poor surface quality, low durability, complex processing steps, and restricted geometric freedom, hindering the repeatable production of thin-walled, continuous structures. Here, an additive manufacturing–facilitated blow molding (AM-BM) approach is introduced, combining the design flexibility of additive manufacturing with the robustness of blow molding. By replacing metal molds with 3D-printed resin ones, AM-BM enables rapid, low-cost fabrication of thin-walled polymeric components with tunable geometry and controllable wall thickness across diverse thermoplastic materials. The thickness control allows thin-walled components to function either as rigid load-bearing elements or as compliant hinges that permit reversible deformation. The versatility of AM-BM is demonstrated through representative examples: multistable structures with geometry-controlled buckling and rich reconfigurability; origami and kirigami structures with extensive design freedom, scalable complexity, and uniform mechanical properties; and soft actuators and robots with ultrahigh load-to-weight ratios, rapid response, and scalable design. Altogether, AM-BM provides an efficient and versatile method for creating thin-walled structures that combine geometric freedom, mechanical functionality, and scalable production.

\end{abstract}

\justifying 
\section{Introduction}

Soft and deformable structures, characterized by their ability to undergo large, reversible deformation, have emerged as a key component for applications requiring adaptability, compliance, and functional versatility~\cite{Nohooji2025-Compliant-Robotics}. They are increasingly exploited in fields ranging from multistable and morphing structures \cite{Pan2019straw, BenAbu2024strawfabrication, Veksler2024}, origami \cite{Melancon2021, Wu2021origami-robotic-arm} and kirigami \cite{Rafsanjani2019kirigami-shell,Rafsanjani2018} to soft robotics \cite{DePascali2022, Chen2021lost-wax}. Such deformability can be achieved either through the use of intrinsically soft materials or through geometric design strategies that reduce structural stiffness. Among the latter, thin-walled structures, characterized by high aspect ratios (between their longitudinal/transverse dimensions and their thickness), offer lightweight yet mechanically robust designs \cite{Liu2024Thin-walled-structures}. Here, we focus on tubular thin-walled structures at laboratory-relevant scales, with lengths and diameters ranging from millimeters to decimeters and wall thicknesses below 500 $\mu$m, which can function either as rigid load-bearing elements or as flexible hinges that bend. 

When a thin-walled structure is made entirely from a homogeneous constituent material, its behavior is fundamentally governed by geometry \cite{Ranzani2018}. Consequently, manufacturing techniques become a critical determinant of its functionality and must be considered carefully throughout the design process. Although thin-walled structures are indispensable and widely used in soft and deformable systems, fabricating functional variants with high aspect ratios remains challenging with existing approaches. Among the most common methods are 3D printing, lost-wax casting, and laser cutting. 

Additive manufacturing (AM) enables the fabrication of nearly arbitrary geometries \cite{Truby2016}. However, existing techniques are inadequate for producing medium-scale (i.e., milli- to decimeter scale) thin-walled structures that combine functionality with large, reversible deformation. In material extrusion–based AM, the resolution is fundamentally limited by the nozzle diameter, making it difficult to print walls thinner than the extrusion width, while interlayer defects often lead to rupture \cite{DePascali2022}. Vat photopolymerization, powder bed fusion, material and binder jetting can directly produce thin-walled structures from soft materials, yet the achievable wall thickness is typically on the same order as the printing resolution \cite{Ligon2017-3D-Print-review, Siddiqui2024-3DP-overview}. As a result, processing defects can induce fragility, reduce durability, and constrain geometric complexity. Moreover, conventional material extrusion and vat photopolymerization methods require support structures during printing, which further complicates the fabrication of compliant, thin-walled geometries \cite{Ngo2018-3DP-review, Diaco2025-vat-photopolymerization}. In contrast, two-photon polymerization, a high-resolution variant of vat photopolymerization, can achieve sub-micron feature sizes and ultra-thin deformable structures, but its small build volume severely limits the overall dimensions that can be produced \cite{Jaiswal2023-2PP}. 

Soft material casting can be used to fabricate thin-walled structures \cite{Shan2015Multistable}, but it is generally limited to relatively simple geometries due to defect formation and challenges associated with demolding thin features \cite{Xu2023-multistable-metamaterials-fabrication, Schmitt2018soft-robot-fabrication}. Lost-wax casting has been developed to produce hollow soft structures with more intricate geometries \cite{Marchese2015lost-wax,Chen2021lost-wax}. Yet, in a laboratory setting the process involves complex, time-consuming procedures and lacks the scalability necessary for efficient batch production.

Laser cutting of thin films can also be employed to fabricate thin-walled origami \cite{Felton2014origami-robot,Wu2021origami-robotic-arm} and kirigami structures \cite{Rafsanjani2019kirigami-shell,Jin2020-multistable-kirigami}. However, origami structures produced using this method require manual folding, making the process labor-intensive and limiting the achievable geometric complexity \cite{Xue2025-origami-fabrication}. Kirigami structures fabricated in this manner are typically restricted to planar configurations, while constructing three-dimensional (3D) shells generally relies on gluing. This assembly process introduces joint regions with double thickness, resulting in local mechanical inhomogeneity.

Blow molding, by contrast, was specifically developed to address the challenge of fabricating hollow thermoplastic structures \cite{Cantor2011blow-molding-introduction, Lee2008blow-molding-introduction}. In this process, a molten polymer is inflated within a mold using pressurized air, allowing it to conform precisely to the mold geometry and enabling the rapid, repeatable production of seamless, thin-walled components. Owing to its scalability and versatility, blow molding has been extensively adopted across applications ranging from small consumer bottles and packaging to large industrial components such as ducts, tanks, and bumpers \cite{Moskatov1968bellow-fabrication, Wagner2014blow-molding-introduction}. However, conventional blow-molding production lines are typically large-scale and capital-intensive. The molds used in these systems are generally machined from metal to ensure dimensional precision and long service life, yet their fabrication demands specialized tooling, extensive machining time, and high costs \cite{Pelin2024drawbacks-traditional-methods}. As a result, such industrial setups are economically viable only for high-volume manufacturing, whereas for laboratory-scale or iterative prototyping they become prohibitively slow and expensive. This cost–time barrier fundamentally limits the rapid design–test iteration cycles essential to research environments and rapid prototyping, motivating the development of fabrication strategies that preserve the advantages of blow molding while introducing greater flexibility and speed.

To this end, we introduce \textit{additive manufacturing–facilitated blow molding} (AM-BM), a hybrid method that combines the scalability of blow molding with the design freedom of additive manufacturing. By replacing conventional metal molds with 3D-printed heat-resistant resin molds, AM-BM greatly shortens fabrication time and lowers costs, while maintaining shape accuracy and re-usability comparable to metal molds, ensuring reliable performance. The required equipment is compact, affordable, and compatible with standard laboratory environments. This approach enables fabrication of polymeric thin-walled structures with consistent mechanical performance, while minimizing material usage and environmental impact. It supports fast design–test iterations, and allows scalable production of customizable components across diverse applications---including multistable structures, origami/kirigami patterns, and soft robotic systems. As we will show, AM-BM not only provides an accessible and sustainable route for producing complex thin-walled geometries. It also enables functionalities such as tunable mechanical responses, precisely engineered arrays of repeating folding and cutting elements, and lightweight yet powerful actuators with enhanced agility---with significant shape and functionality advantages over existing alternatives. Collectively, these advantages establish AM-BM as a unified fabrication framework that integrates functionality, adaptability, and manufacturability. AM-BM offers a practical platform for research, education, and versatile production, enabling the exploration of fundamental design principles and mechanical performance as well as rapid prototyping of functional thin-walled architectures.

\section{Results and discussion}
\subsection{Overview of additive manufacturing-facilitated blow molding}

\begin{figure}
\centering
  \includegraphics[scale=1]{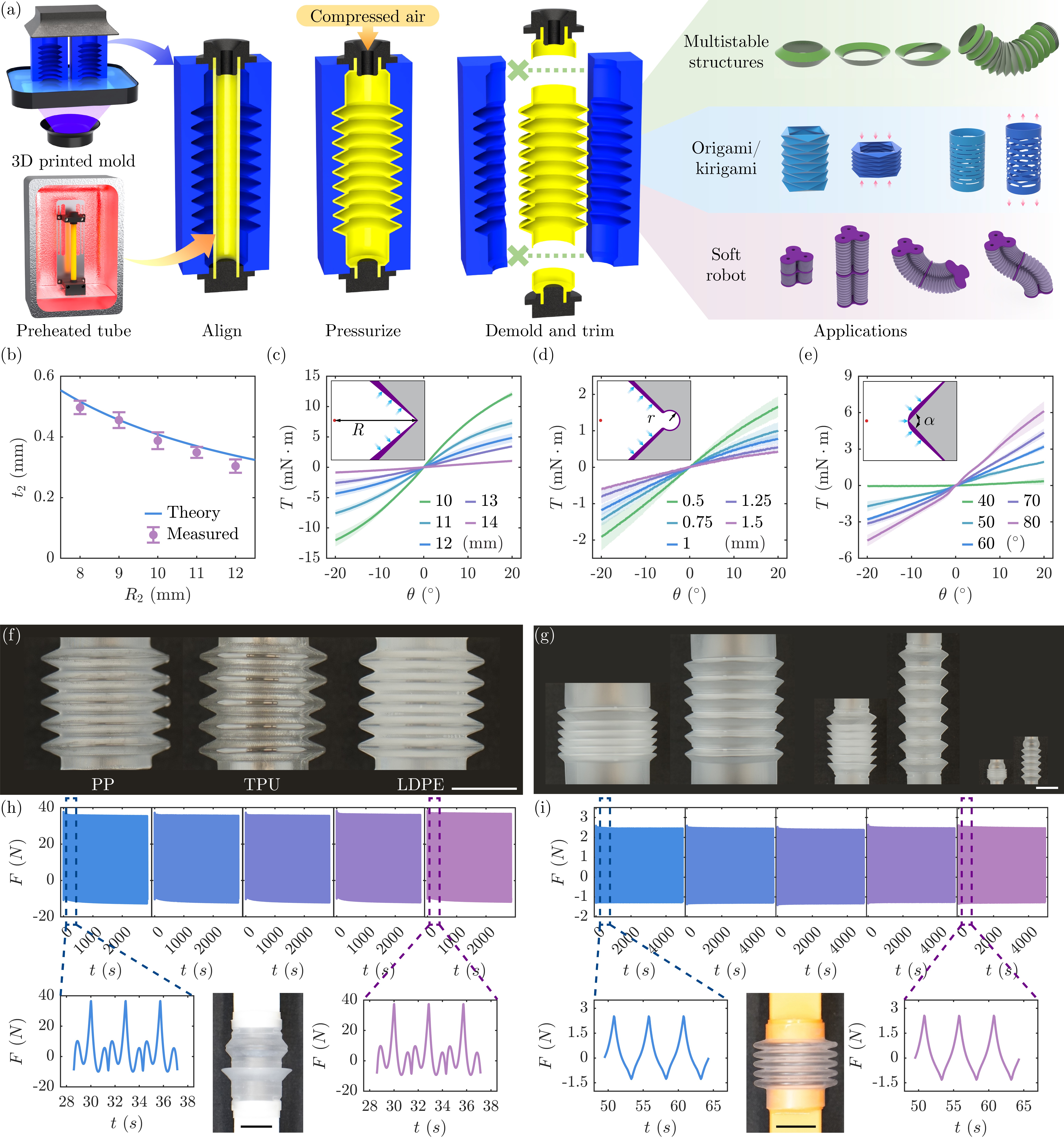}
  \caption{\textbf{Fabrication process and demonstration of AM-BM}.  (a) Schematic illustration of the AM-BM fabrication process and potential applications. (b) Relation between mold radius $R_2$ and resulting wall thickness $t_2$. (c-e) Measured torque response of (c) convex hinges with different vertex radii, (d) convex hinges with different arc radii, and (e) concave hinges with different angles (the inset schematics are the cross-section views of blow-molded hinges, the 3D view is shown in Figure~S1, Supporting Information, the red dots denote the central axes, the purple regions represent the expanded thermoplastic material, and the gray regions indicate the mold; shaded envelopes in the data plots show standard deviations of 3 tests). (f) Bellows fabricated from different thermoplastic materials (scale bar: 10~mm). (g) Multistable structures of various sizes (scale bar: 10~mm). (h,i) Cyclic durability tests of (h) a multistable structure and (i) a bellow (scale bar: 10~mm). }
  \label{fig1}
\end{figure}

The fabrication workflow of the AM-BM process is illustrated in Figure~\ref{fig1}a and Video S1 (Supporting Information). First, the molds are produced using a vat-photopolymerization 3D printer with heat-resistant resin. Next, the parison---a segment of a thermoplastic tube---is preheated in an oven to its glass transition temperature. After alignment of the parison with the mold, the mold halves are closed, and compressed air is applied to the interior of the parison. The resulting internal pressure drives radial expansion, forcing the parison to conform to the mold contours. The formed part is subsequently cooled, demolded, and trimmed to obtain its final geometry.

The wall thickness of thin-walled structures plays a critical role in determining their mechanical behavior \cite{Wagner2020influence-of-thickness}. Therefore, achieving precise thickness control is a key design consideration. As a simple, instructive benchmark, we analyze a uniform tubular parison in a cylindrical mold whose walls are parallel to the parison's axis of symmetry. Assuming volume conservation during the forming process and a constant mass density, the cross-sectional area of the parison remains constant before and after pressurization \cite{Galuppo2025volume-conservation}. This yields a geometric relation between the initial parison and the final mold geometry:
\begin{equation}
\pi R_1^2 - \pi (R_1 - t_1)^2 = \pi R_2^2 - \pi (R_2 - t_2)^2,
\end{equation}
where $R_1$ and $R_2$ are the outer radius of the parison and inner radius of the mold, respectively, while $t_1$ and $t_2$ are the wall thicknesses before and after pressurization, respectively. Rearranging leads to the resulting thickness $t_2$ of the formed thin-walled structure as
\begin{equation}
\label{thickness equation}
t_2 = R_2 - \sqrt{R_2^2 + t_1^2 - 2R_1t_1}.
\end{equation}

For validation, we used AM-BM to produce five cylindrical tubes with uniform radii ranging from 8~to 12~mm, from a parison with a fixed outer radius of 4.5~mm and wall thickness 1~mm. The experimental results in Figure \ref{fig1}b align well with Eq.~\ref{thickness equation}, showing that, for a constant parison geometry, increasing the mold radius results in a corresponding decrease in the final wall thickness of the blow-molded structure.

To enable reversible large deformations and enhance the functionality of blow-molded thin-walled components, we introduce a design featuring localized hinge regions. These hinges are designed to be thinner than their adjacent relatively rigid sections, allowing for significant relative motion between connected segments without compromising the structural integrity of the overall part---of importance, e.g., for multistable structures, origami and bellows. The design of such hinges requires considering convex and concave configurations. In a convex hinge (Figure \ref{fig1}c,d), the local radius at the hinge is larger than that of the adjacent regions, whereas in a concave hinge (Figure \ref{fig1}e) the local radius is smaller.

For convex hinges, Eq.~\ref{thickness equation} indicates that a larger mold radius results in a thinner wall. Leveraging this principle, we designed specimens in which the edge radius is intentionally larger than the radii along the adjoining planes, ensuring the wall is thinnest at the hinge location. To test if such local radius variations are sufficient to enable hinge-like behavior without additional structural modifications, we fabricated convex hinge specimens with different vertex radii, each having a $90^\circ$ angle (Figure \ref{fig1}c). During experiments, one side of the hinge was fixed to a torque sensor, while the other side was connected to a motor (Figure S1, Supporting Information). As expected, convex hinges with larger radii exhibited lower actuation torques due to their reduced thickness. Thus, for sufficiently large radii, no additional geometrical modifications are needed to create an effective hinge.

When the vertex radius is too small, the resulting wall thickness at the hinge is insufficiently reduced to allow smooth rotational motion. As a remedy, we incorporated a secondary arc at the hinge (Figure \ref{fig1}d), increasing the local perimeter and thereby reducing the wall thickness according to the volume conservation principle. We fabricated a series of specimens with $90^\circ$ bends and a vertex radius of 10 mm. The arc radius $r$ was varied from 0.5~mm to 1.5~mm, and the required actuation torque using the same experimental setup as before. Results confirm that enlarging the arc radius decreases the rotational torque, demonstrating improved hinge performance. However, when $r > 1~\text{mm}$, the hinge becomes mechanically unstable: instead of enabling pure rotation about the vertex, the structure tends to deform in additional directions, introducing unintended degrees of freedom.

Concave hinges, by contrast, benefit from a distinct mechanism during blow molding: when the mold features an inward-facing acute angle, the applied air pressure during blow molding drives the molten material outward to the adjoining surfaces. Due to the material’s fluidity at this stage, it is redistributed away from the vertex, leaving a locally thinner wall at the hinge. To investigate this phenomenon, concave hinges with vertex radius of 7~mm and varying opening angles $\alpha$ were fabricated to measure their rotational stiffness (Figure \ref{fig1}e). The results show that smaller angles (i.e., sharper concave corners) lead to reduced rotational resistance, confirming the formation of flexible hinge regions. Only when $\alpha < 40^\circ$, the thinning becomes excessive and the resulting hinge loses mechanical durability.

By controlling the wall thickness and hinge geometry, this general hinge design strategy unlocks new functional capabilities and deformation modes in thin-walled structures, which we will demonstrated in multistable structures, origami/kirigami, and soft robots.

AM-BM is compatible with a wide range of thermoplastic polymers, including polypropylene (PP), thermoplastic polyurethane (TPU), and low-density polyethylene (LDPE), which exhibit significant differences in physical properties such as elastic modulus, viscoelastic response, and glass transition temperature \cite{VanKrevelen2009Properties-of-Polymers}. Figure \ref{fig1}f shows bellows fabricated from those three materials, illustrating the versatility of AM-BM. Thus, combined with the capabilities of additive manufacturing, our method enables rapid production of complex geometries of diverse geometries, dimensions and mechanical properties. Figure \ref{fig1}g shows AM-BM-fabricated multistable units with diameters ranging from 9~mm to 50~mm. Regardless of their size or shape, all units exhibit multistability.

Next, we evaluated the durability of two types of samples, a multistable single-cell unit and a monostable bellow, by subjecting them to cyclic uniaxial extension and compression. Each durability test consisted of 5 sets of 1000 load cycles (Figure \ref{fig1}h,i). Owing to the viscoelastic nature of PP (Figure S2, Supporting Information), gradual changes in the mechanical response were observed within each test group. However, after approximately one day of relaxation, we conducted a second set of 1000 cycles and found that the samples exhibited similar behavior to the first test group. This suggests that no permanent damage had occurred. Both structures endured five such test groups (a total of 5000 cycles) without significant degradation in performance, demonstrating the robustness of AM-BM-fabricated components under repeated deformation.

\subsection{Multistable structure}

\begin{figure}
  \centering
  \includegraphics[scale=1]{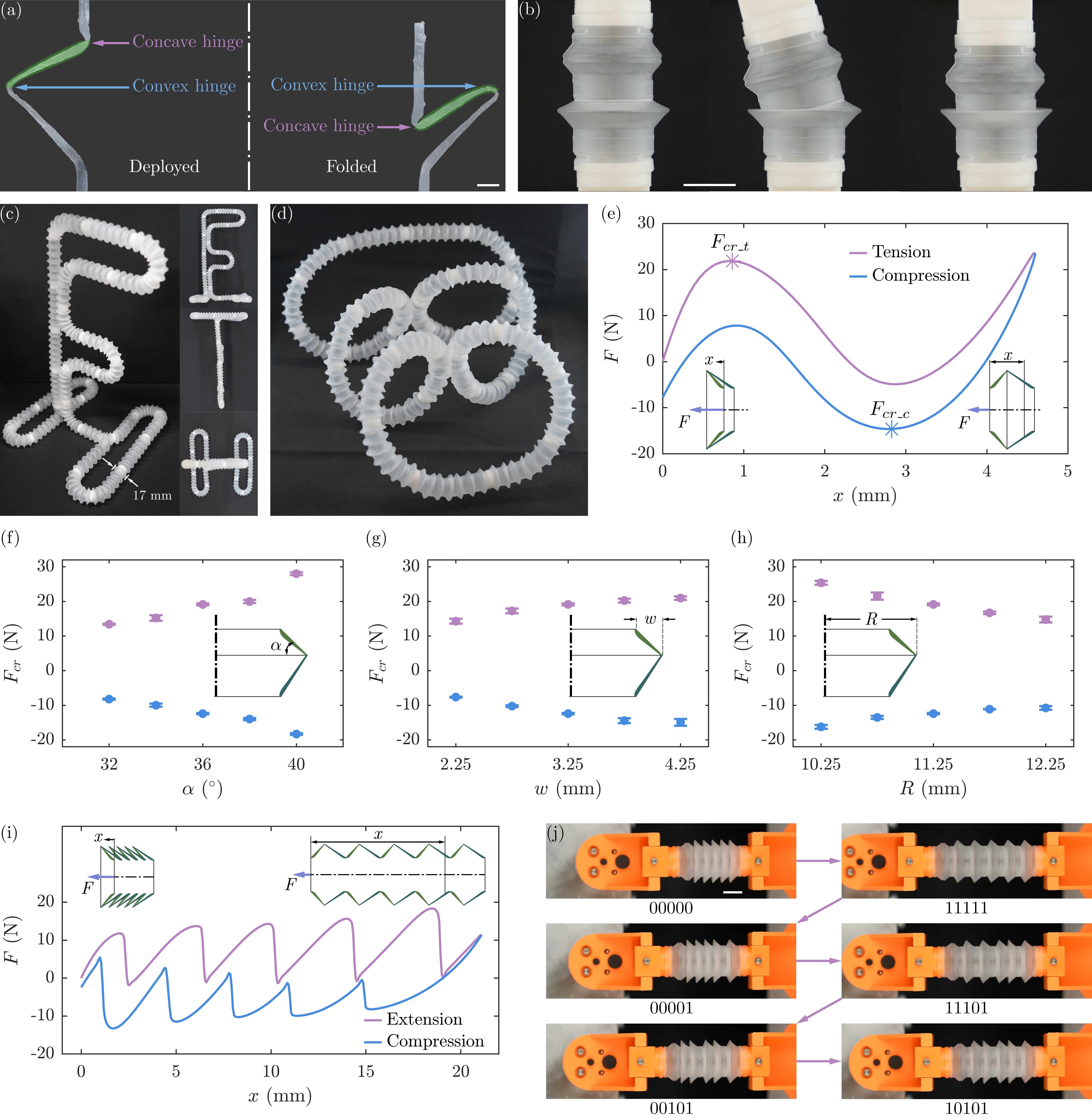}
  \caption{\textbf{Multistable structures}. (a) Micrographs of the cross-section of a multistable conical shell in the deployed (left) and folded (right) configurations (scale bar: 1~mm). (b) Multistable structure in the deployed (left), folded (middle), and bent (right) stable configurations. (scale bar: 10~mm) (c,d) Reconfiguration of a multistable structure into (c) a 3D ``ETH'' and (d) a double-helix shape. (e) Force–displacement response of a single multistable unit cell. (f–h) Variation of the critical buckling forces $F_\text{cr,t/c}$ with design parameters of the conical shell: (f) tilting angle~$\alpha$, (g) width~$w$, and (h) outer radius~$R$. (i) Force–displacement relation of a straw made of five multistable unit cells with different critical buckling forces connected in series. (j) Sequential reconfiguration of the multistable structure from the “00000” to “10101” state (scale bar: 10~mm; see also Video S2, Supporting Information). }
  \label{fig2}
\end{figure}

Inspired by bendy drinking straws, multistable structures composed of serially connected conical shells (also referred to as frusta) have attracted growing attention, owing to their ability to undergo large, reversible deformations and exhibit multiple stable configurations. These structures have been explored in applications such as impact mitigation \cite{Pan2019straw}, reconfigurable antennae \cite{Liu2022}, shape-morphing systems \cite{Veksler2024}, and metafluids with multistable thermodynamic behavior \cite{Peretz2022}. However, most studies rely on commercially available bendy straws fabricated by conventional blow molding, for which the geometry, dimensions, and mechanical properties are fixed. Attempts to reproduce or customize straws through casting or material-extrusion additive manufacturing \cite{Bende2018casting-3D-printing-straw} have been hindered by limited precision and structural defects. Although a die-forming method was recently introduced for customizable fabrication \cite{BenAbu2024strawfabrication}, manual operation still prevents achieving uniform quality and scalability. Overcoming these limitations, AM-BM allows fabricating multistable structures composed of series of frusta, enabling precise geometric control, scalable production, and consistent product quality for consistent mechanical behavior.

The cross-section of a representative multistable unit cell—consisting of a pair of connected conical shells—is shown in Figure~\ref{fig2}a and Figure~S3 (Supporting Information). The upper frustum, capable of large reversible deformation, is highlighted in green. It is connected to the neighboring relatively stiff frusta through a concave hinge at the top and a convex hinge at the bottom. Consequently, the upper frustum can stably adopt both a folded and a deployed configuration.

As illustrated in Figure~\ref{fig2}b, the unit cell has three distinct stable configurations: 1.~the frustum is deployed; 2.~the frustum is folded; 3.~the frustum is deployed on one side but folded on the other. Because of the structure’s axisymmetry, the third configuration corresponds to an infinite number of stable states. By connecting multiple such multistable unit cells in series, a wide range of complex and reconfigurable straw architectures can be achieved.
For example, Figure~\ref{fig2}c shows a straw composed of approximately 200 multistable unit cells. In one of its stable configurations, the 3D structure forms the letters ``E'', ``T'', and ``H'' when viewed from the front, left, and top, respectively. Owing to the large number of interconnected multistable units, the same structure can be reconfigured into a completely different shape; e.g., a double helix, as shown in Figure~\ref{fig2}d.

AM-BM allows us to readily customize the design parameters of multistable straws and investigate their influence on the mechanical response. First, displacement-controlled uniaxial extension and compression tests were performed on a single unit (Figure~\ref{fig2}e). During extension, the test starts from the folded stable state, passes through the deployed stable state, and continues stretching beyond. The extension curve, shown in purple, terminates before any plastic deformation occurs. The limit point during extension is marked with a purple star. The force at this point is the tensile critical buckling force~$F_\text{cr\_t}$. After reaching the maximum displacement, the structure was compressed back to its initial configuration, following the blue curve. The corresponding limit point during compression, marked by a blue star, is defined as the compressive critical buckling force~$F_\text{cr\_c}$. Due to the viscoelastic nature of the material, the extension and compression curves form a hysteresis loop (which is not captured by the theoretical model \cite{Ilssar2022}, since the latter neglects inelastic effects).

To explore how design parameters influence the mechanical response and to demonstrate how AM-BM easily enables such design variations, we manufactured straws with three  systematically varied key geometric parameters. We first examined the effect of the inclination angle $\alpha$ between the deformable frustum and the horizontal plane (Figure~\ref{fig2}f). Increasing~$\alpha$ leads to larger absolute values of both positive and negative critical buckling forces, indicating that the frustum exhibits a higher buckling load. The second parameter is the frustum width~$w$, defined as the difference between the outer and inner radii. With the outer radius $R$ and inclination angle $\alpha$ fixed, a wider frustum exhibits higher critical buckling forces (Figure~\ref{fig2}g). Finally, the outer diameter $R$ was varied, while maintaining a constant angle~$\alpha$ and width~$w$. Increasing $R$ results in a reduction of the absolute critical buckling forces (Figure~\ref{fig2}h). Overall, the experimental observations show good qualitative agreement with a theoretical straw model \cite{Ilssar2022} (Figure S4, Supporting Information)---even though the latter relies on several idealizations (including infinitely thin creases connecting adjacent conical shells, linear elastic material behavior, and uniform wall thickness).

Capitalizing on the multistability of a single unit, multiple conical shells can be connected in series to form a multistable structure with a prescribed buckling sequence. As an example, we combined five unit cells and varied the inclination angle~$\alpha$ from 32$^\circ$ to 36$^\circ$, such that the critical buckling force increases from unit cell to unit cell. Since the units are mechanically connected in series, uniaxial actuation imposes approximately the same force on every unit cell (given that the actuation is sufficiently slow, so the static forces dominate). Consequently, during both tensile and compressive loading, the buckling sequence proceeds from left to right (Figure~\ref{fig2}i).

Remarkably, this spatially graded straw can reconfigure among a total of $2^5 = 32$ stable equilibria using only a single actuator, which showcases the potential of this highly underactuated system for mechanical information encoding and storage. For brevity, we denote the fully deployed and fully folded states of a single unit cell as ``1'' and ``0'', respectively. Figure~\ref{fig2}j and video~S2 (Supporting Information) illustrate an example in which the straw is reconfigured from ``00000'' to ``10101'' in five sequential steps: 1.~stretch to the fully deployed state ``11111''; 2.~compress until the first four unit cells collapse, yielding ``00001''; 3.~stretch so that the first three unit cells deploy again, producing ``11101''; 4.~compress so that the first two units fold, giving ``00101''; 5.~stretch once more to deploy the first unit, reaching the target state ``10101''. Given the design freedom enabled by AM-BM and the structure-property relations established here, the attainable configuration space of multistable straw architectures is vast, with geometry defining the potential energy landscape and simple actuation enabling complex programmable transitions among its stable states.

\subsection{Tubular Origami and Kirigami}

\begin{figure}
    \centering
  \includegraphics[scale=1]{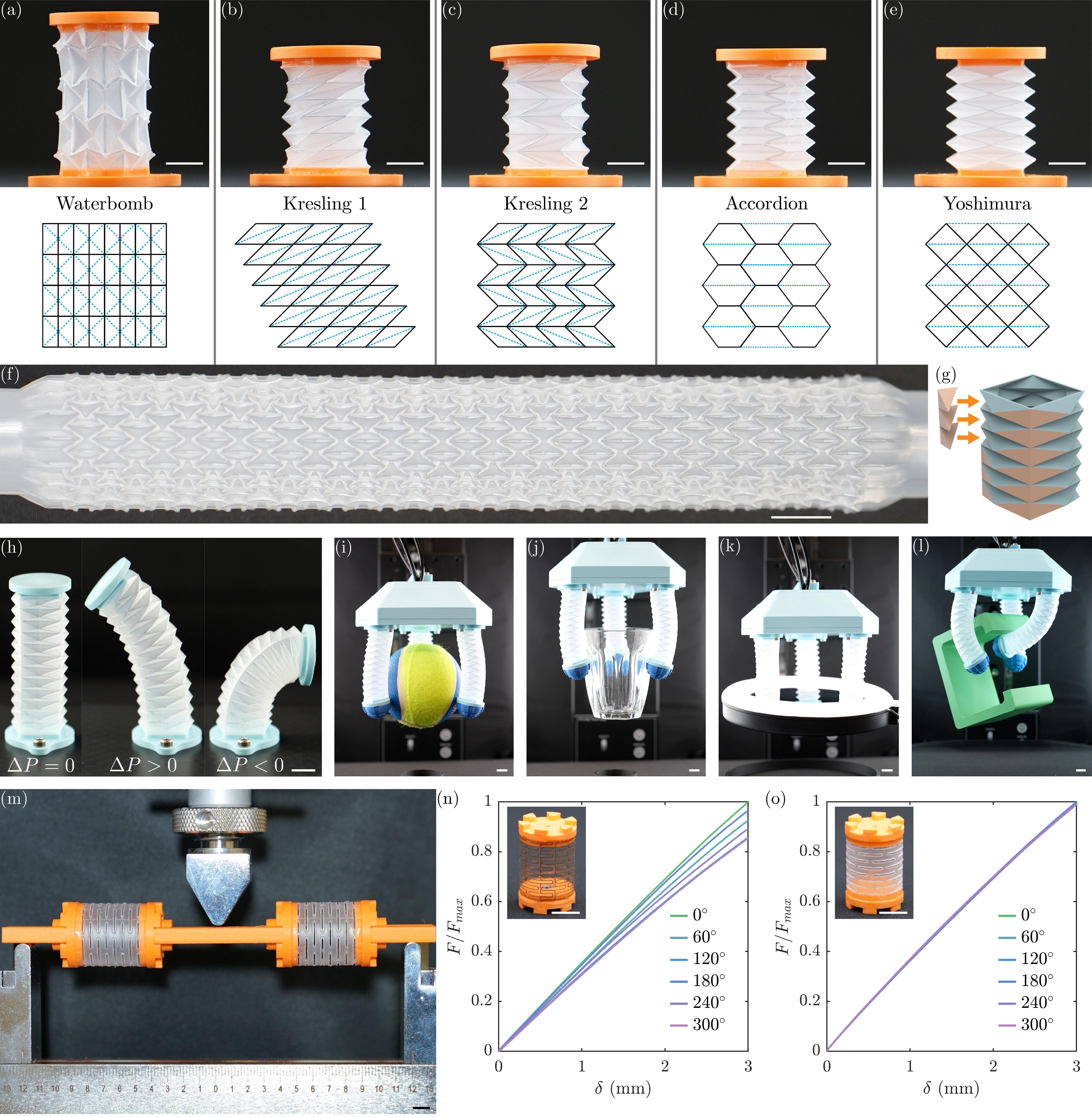}
  \caption{\textbf{Origami and kirigami} (scale bars: 10~mm). (a–e) Designs of different origami and kirigami patterns, where black lines denote mountain folds and blue lines denote valley folds: (a) Waterbomb, (b) Kresling 1, (c) Kresling 2, (d) Accordion, and (e) Yoshimura. (f) A waterbomb origami with 496 unit cells. (g) Modified Yoshimura origami design. (h) Pneumatic origami actuator (based on the adapted Yoshimura design) under zero (left), positive (middle), and negative (right) gauge pressure $\Delta P$. (i–l) Origami gripper grasping different objects: (i) a ball, (j) a glass, (k) a ring light, and (l) a ``C''-shaped object. (m) Three-point bending test setup for the tubular kirigami structures. (n,o) Normalized three-point bending test results at different orientations for (n) laser-cut kirigami and (o) AM-BM-fabricated kirigami.}
  \label{fig3}
\end{figure}

Origami and kirigami, though rooted in artistic expression, have become powerful design frameworks for achieving complex shape morphing, programmable mechanical responses, and multifunctional behavior \cite{Misseroni2024Origami-overview, Jin2024Kirigami_overview}. Origami exploits crease patterns to realize deployable mechanisms and reconfigurable architectures, while kirigami introduces patterned cuts that enable large stretchability and rich three-dimensional transformations. These approaches underpin applications in robotics \cite{Felton2014origami-robot, Rafsanjani2018, Wu2021origami-robotic-arm}, flexible electronics \cite{Cho2014stretchable-electronics,Jiang2022stretchable-electronics}, biomedical devices \cite{Meng2022kirigami-medical, Suzuki2020origami-biomedical}, architected materials \cite{Jin2020-multistable-kirigami,Zhao2025origami-metamaterials}, and aerospace structures \cite{Zirbel2013origami-aerospace, Hussain2023origami-aerospace}. However, despite their geometric versatility, the scalable fabrication of thin-walled tubular origami and kirigami structures with uniform mechanical properties remains challenging using conventional methods \cite{Sharma2025tubular-origami-overview}. This motivates the use of AM-BM with the hinge design principles introduced above as a reliable and customizable route to produce tubular origami and kirigami. Let us demonstrate this first for origami.

Origami structures are modeled as facets connected by hinges, whose folding pattern defines their overall kinematics. Within the blow-molding framework, the flat regions of the mold form the plates, while the mountain and valley folds are created through the outer and inner hinge geometries, respectively.

Figure~\ref{fig3}a-e presents five representative origami tessellation patterns: Waterbomb, Kresling~1, Kresling~2, Accordion, and Yoshimura \cite{Sharma2025tubular-origami-overview}, all fabricated by AM-BM and each exhibiting distinct folding geometries and mechanical behavior (video S3, Supporting Information). The Waterbomb pattern shows an auxetic response, with the tube narrowing laterally at its midsection under axial compression \cite{Ma2020Waterbomb}. The Kresling patterns exhibit coupled axial–rotational motion: in Kresling~1 identically oriented units lead to cumulative rotation and pronounced twisting during loading, whereas in Kresling~2 alternating unit orientations cancel rotation, yielding purely axial deformation without twist \cite{Zang2024Kresling}. The Accordion and Yoshimura patterns provide large axial compliance and can accommodate substantial bending in multiple directions \cite{Wickeler2023-4D-printed-origami}. To demonstrate scalability, Figure~\ref{fig3}f shows a Waterbomb origami comprising 496 square unit cells. Despite the much higher cell count, the design and fabrication effort remains comparable to that of the 24-cell structure in Figure~\ref{fig3}a. Collectively, these examples highlight AM-BM’s geometric fidelity, seamless hinge integration, hermetic sealing, and highly scalable design-to-fabrication workflow.

The AM-BM approach enables the fabrication of both conventional origami designs, which can be formed by folding a single continuous sheet, and more complex patterns that require a combination of folding and cutting. Figure~\ref{fig3}g illustrates an example in which two columns of tetrahedral units (with orange color) are incorporated into a Yoshimura pattern---such a configuration cannot be realized by folding a single continuous sheet. In contrast, AM-BM can produce both conventional fold-only patterns and more complex cut-and-fold geometries with equal ease, without the need for additional fabrication steps. In the modified Yoshimura pattern, the added tetrahedral columns restrict axial extension, while the opposite side remains deformable, allowing the structure to bend within a fixed plane under positive and negative gauge pressure (Figure \ref{fig3}h). Leveraging this directional bending capability, the adapted origami design functions effectively as a pneumatic origami gripper (Figure~\ref{fig3}i-l, video S4, Supporting Information). The gripper can manipulate objects of various shapes and sizes. For the ball and the glass (Figure~\ref{fig3}i,j), negative gauge pressure is first applied to bend the origami fingers outward, opening the gripper. Upon approaching the object, positive gauge pressure is applied, causing the fingers to bend inward and securely grasp. For the annular object in Figure~\ref{fig3}k, the gripper begins in a closed state; when negative gauge pressure is applied after reaching the object, the fingers bend outward to hold the ring from the inside. For asymmetric objects, such as the ``C''-shaped profile in Figure~\ref{fig3}l, a non-symmetric actuation strategy can be used. In this specific case, two fingers first bend outward to approach the object, followed by inward bending of the third finger to tilt it into position. Finally, reversing the gauge pressure in the first two fingers from negative to positive completes the grasping process.

The same AM-BM strategy can be extended to directly fabricate kirigami on tubular thin-walled structures. So far, tubular kirigami has typically been produced by laser cutting a flat plastic sheet and subsequently gluing opposite edges to form a tubular structure \cite{Rafsanjani2018, Rafsanjani2019kirigami-shell}. This process creates a joint region with double thickness (Figure~S5, Supporting Information), leading to local mechanical inhomogeneity. In our AM-BM approach, grooves are incorporated into the mold at the intended kirigami cut locations. During blow molding, the applied air pressure pushes the molten material into these grooves, forming small protrusions. After molding, the outer surface is polished to remove the excess material, leaving behind the slits that define the kirigami pattern (Figure~S6, Supporting Information). To evaluate the mechanical uniformity of the resulting structures, we AM-BM-fabricated tubular kirigami samples with sixfold symmetry consisting of evenly spaced horizontal slits. For comparison, reference samples with identical geometric parameters were produced using the conventional laser-cut-and-glue method. Three-point bending tests (Figure \ref{fig3}m) were performed in six differently rotated orientations of the krigami tube (from 0$^\circ$ to 300$^\circ$ in 60$^\circ$ increments) corresponding to the six-fold symmetry of the structure. The bending stiffness of the laser-cut samples varied by more than 14\% (Figure~\ref{fig3}n), whereas the AM-BM-fabricated samples exhibited a maximum devitation of 1.2\% (Figure~\ref{fig3}o). This demonstrates that AM-BM significantly enhances the mechanical uniformity of tubular kirigami structures.

\subsection{Soft robotic actuators}

\begin{figure}
    \centering
  \includegraphics[scale=1]{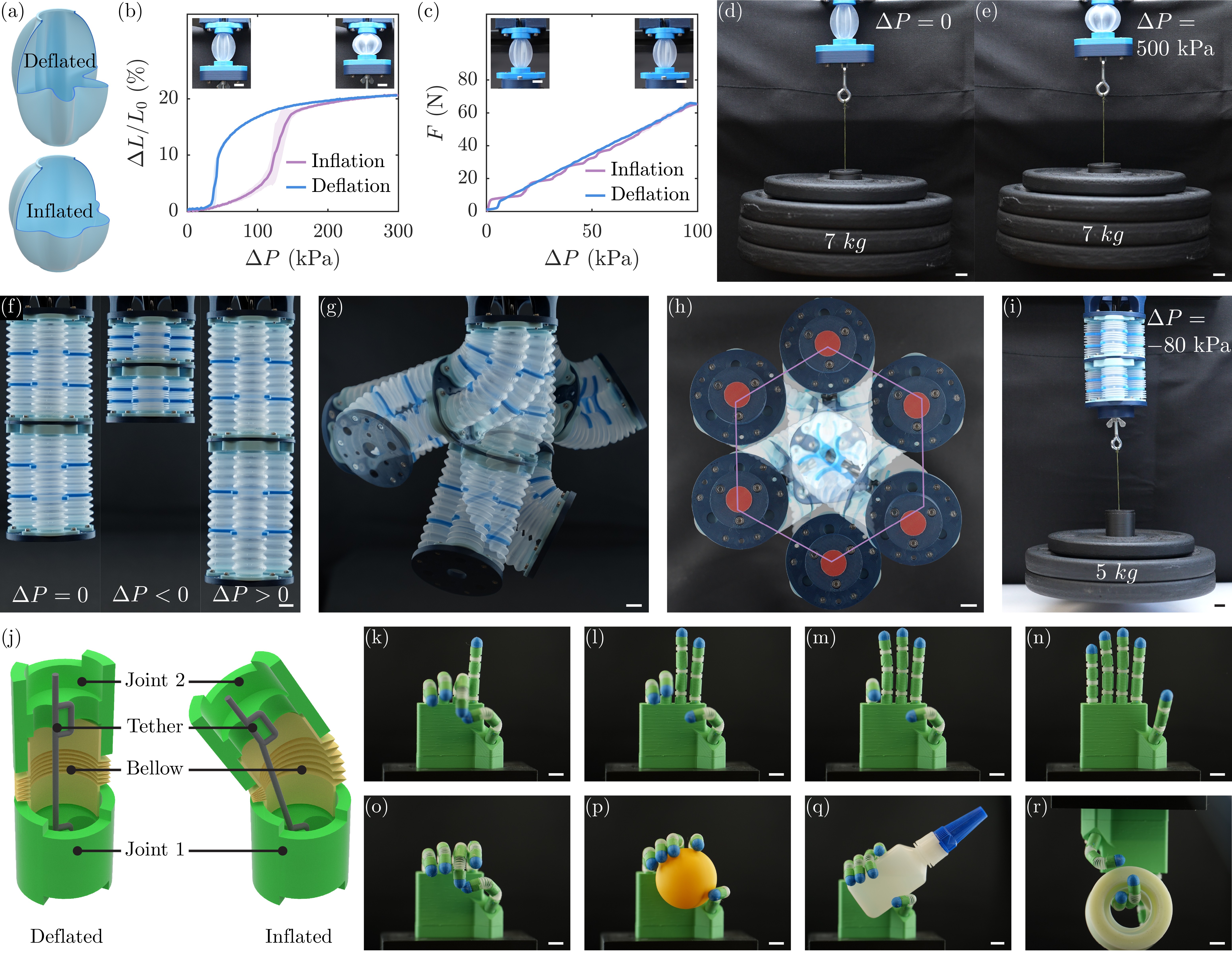}
  \caption{\textbf{Soft robotic systems fabricated using AM-BM} (scale bars: 10~mm). (a–e) Pneumatic artificial muscle. (a) Working principle of the artificial muscle. (b) Strain–pressure relation under a 1~kg load. (c) Force–pressure relationship with both ends fixed. (d) Unactuated state supporting a 7~kg payload. (e) Actuated state with a 7~kg payload. (f–i) Soft robotic arm. (f) Actuation under zero (left), negative (middle), and positive (right) gauge pressure $\Delta P$. (g) Demonstration of the arm’s workspace. (h) Hexagonal trajectory achieved through combined bending and rotation. (i) Lifting of a 5~kg load. (j–r) Bionic robotic hand. (j) Design of an individual bending joint. (k–o) Representative gestures including (k) number ``1", (l) number ``2", (m) number ``3", (n) fully open palm, and (o) closed fist. (p–r) Grasping different objects, including (p) a ball, (q) a bottle, and (r) a roll of tape.}
  \label{fig4}
\end{figure}

Soft robotics, grounded in compliant materials and geometrically engineered flexibility, has emerged as a powerful paradigm for creating systems capable of safe interaction with the environment, large deformations, and adaptive motion \cite{Rus2015soft-robotics-overview, Whitesides2018soft-robotics-overview}. Soft robotic technologies span a wide range of applications, including manipulation and grasping \cite{Terryn2017soft-gripper, Liu2020softrobot-manupulation}, biomedical devices \cite{Kim2019softrobot-biomedical, Wang2023softrobot-biomedical}, smart wearables \cite{Zhou2024softrobot-wearable, Clode2024softwearable}, and bio-inspired locomotion \cite{Katzschmann2018softrobot-fish, Ren2019softrobot-locomotion}. However, fabricating soft actuators with high force-to-weight ratios, reproducible mechanical behavior, and scalable manufacturability remains a challenge \cite{Xavier2022pneumatic-actuator-overview}. This calls for fabrication strategies capable of producing lightweight, high-strength, and robust soft actuators. AM-BM offers the potential to meet these needs by enabling geometrically precise and mechanically reliable soft robotic components.

Previous work has investigated the use of conventional additive manufacturing techniques for fabricating pneumatic artificial muscles \cite{DePascali2022}. However, the maximum output force is constrained by the weak interlayer bonding intrinsic to layer-by-layer additive manufacturing methods, which introduces structural discontinuities and reduces overall mechanical robustness. In contrast, AM-BM produces monolithic, one-piece structures without interlayer interfaces, thereby eliminating this limitation and enabling actuators with significantly higher load-to-weight ratios. Moreover, AM-BM requires no supporting structure, thereby simplifying post-processing and yielding smooth, defect-free outer surfaces. Following the same actuator design as in \cite{DePascali2022}, we fabricated a pleated membrane structure with a cross-section composed of a closed series of elliptical arcs (Figure~\ref{fig4}a). Upon pressurization, the actuator expands radially and contracts longitudinally. In the loaded-actuation test, with the top end fixed, the actuator achieved a 23\% contraction while lifting a 1~kg load at an input gauge pressure of 300~kPa (Figure~\ref{fig4}b). In the blocked-force test, both ends were fixed and the actuator underwent five pressurization cycles; the peak contraction force reached 60~N at maximum gauge pressure input of 100~kPa (Figure~\ref{fig4}c). Furthermore, this actuator, weighing only 0.7~g, was able to lift a 7~kg weight with gauge pressure of 500~kPa (Figure~\ref{fig4}d,e, Video S5, Supporting Information), corresponding to a stroke of 18.7\% and an exceptional load-to-weight ratio of approximately~10,000---one to two orders of magnitude higher than comparable directly 3D-printed actuators  \cite{DePascali2022}. These results demonstrate that AM-BM not only enables seamless and lightweight fabrication but also dramatically enhances the mechanical robustness and force output of soft actuators.

The demonstrated capability of AM-BM to produce high-performance soft actuators further extends to the development of a modularized soft robotic arm with blow-molded bellows. The arm in Figure~\ref{fig4}f consists of two stacked segments, each containing three individual bellow chambers. Within each segment, three bellows are connected in parallel and mechanically supported by two intermediate plates positioned between the top and bottom fixtures. As each bellow operates as an independent module, any damaged component can be readily replaced without affecting the integrity of the overall system. The robotic arm achieves a large range of motion through the coordinated contraction and elongation of its bellows. Figure~\ref{fig4}f shows its overall length under $0$~kPa, $-50$~kPa, and $20$~kPa gauge pressure. Figure~\ref{fig4}g shows several attainable configurations, demonstrating high agility and a large workspace. During open loop cyclic actuation, the arm underwent coordinated bending that follows a hexagonal trajectory (Figure~\ref{fig4}h). Benefiting from the low mass and inertia of the thin-walled bellows, the robotic arm achieves rapid motion. Video S6 (Supporting Information) shows the arm performing continuous circular motion at a frequency of 5~Hz. The load-bearing capability of the robotic arm was also evaluated. Despite a total weight of only $\sim$150~g, the arm can lift a~5~kg load under a gauge pressure of~$-80$~kPa (Figure~\ref{fig4}i), demonstrating the exceptional strength-to-weight ratio and structural robustness achieved through the AM-BM fabrication method.

Building on the same AM-BM-fabricated bellows used in the soft robotic arm, we further demonstrated the scalability of this approach by developing a compact five-finger bionic hand. While the rigid segments were produced via FDM printing, the joints were fabricated using miniature AM-BM bellows with a diameter of~8~mm. In this design, each joint is intended to bend without axial contraction. However, the as-fabricated bellows are initially compressible. To overcome this, the thermoplastic bellows were fixed in their fully compressed state and subjected to a brief heat treatment using a hot-air gun (Figure~S7, Supporting Information). After cooling, the bellows retained the compressed geometry even without external constraint, yet remained capable of elongation when pressurized. This simple post-processing step serves as an effective complement to AM-BM, enabling permanent adjustment of the initial shape and mechanical properties, thereby extending the method’s overall versatility in structural design and functionality. With this fabrication and post-processing strategy, a bionic hand was assembled to illustrate the versatility of blow-molded bellows in a compact, multi-joint system. Each joint provides a single bending degree of freedom. To constrain the bending direction, an eccentric string constraint was added so that bending occurs only within the plane defined by the string and the bellow’s symmetry axis (Figure~\ref{fig4}j). Two or three bellows were combined with rigid connecting segments to form an individual finger, and five such fingers were mounted onto a rigid palm. Each finger is independently actuated by positive or negative air pressure. The assembled bionic hand can perform simple gestures such as counting (``1'', ``2'', ``3''), fully opening the palm, or making a fist (Figure~\ref{fig4}k-o). Moreover, it can grasp various objects, demonstrating both dexterity and adaptability (Figure~\ref{fig4}p-r).

\section{Conclusion}

In summary, we have developed an effective fabrication method---AM-BM---for producing thin-walled structures, which offers repeatability, durability, cost efficiency, short processing time, and scalability for batch production. This method allows tuning of geometry and wall thickness while producing seamless and mechanically robust structures across diverse thermoplastic material.
The versatility of AM-BM has been demonstrated through its application in three representative domains, in combination with a flexible hinge design. In multistable structures, AM-BM enables the fabrication of reconfigurable architectures with tunable stability and mechanical responses. The resulting highly underactuated systems exhibit potential for mechanical information encoding and storage as well as for shape morphing deployable structures. In origami and kirigami, AM-BM enables the realization of intricate folding and cutting patterns with distinct kinematic responses, uniform mechanical properties, and scalable geometric complexity. While increased geometric complexity may require more elaborate mold designs, the blow-molding step itself proceeds with similar processing time once the mold is prepared, enabling highly complex architectures to be formed with comparable fabrication effort to simpler counterparts. Looking forward, such AM-BM-enabled origami and kirigami architectures may provide new design pathways for adaptive structures, morphing devices, and multifunctional mechanical systems. In soft robotic systems, AM-BM yields lightweight yet robust actuators and assemblies, offering enhanced load-to-weight ratios, dexterity, adaptability, and rapid response, forming a versatile basis for the design of efficient and high-performance soft robotic systems.
Altogether, AM-BM provides a unified, efficient, and versatile route to design and manufacture a broad range of functional thin-walled structures, bridging the gap between geometric freedom, mechanical functionality, and scalable production.

\section{Experimental Section}
\threesubsection{Fabrication}\\
For the AM-BM process, molds were fabricated by digital light synthesis (DLS) on a Carbon M2 3D printer with a layer thickness of 25$\mu$m. A high-temperature resistant resin (Loctite 3D IND147) was used as the printing material. After printing, parts were immersed in isopropyl alcohol (IPA) on an orbital shaker for 5~minutes to remove uncured resin, followed by 30~minutes of post-curing in a UV chamber to ensure complete cross-linking. During the blow molding process, a gauge pressure of 100~kPa was used. The connectors for the blow-molded components were fabricated from polylactic acid (PLA) on a Bambulab X1 Carbon FDM printer. The connection between the blow-molded PP or LDPE parts and the PLA connectors was achieved by Loctite~770 polyolefine primer and Loctite~406 instant adhesive.

\threesubsection{Experimental characterization}\\
Photos and videos were recorded by a Sony Alpha 7S III camera. Mechanical characterization, including rotational stiffness tests of hinges, uniaxial tension and compression of multistable unit cells, cyclic tests of multistable structures and bellows, and three-point bending experiments, was performed using an Instron 5943 universal testing machine. Rotational stiffness measurements were carried out at an angular displacement rate of $180^\circ/\mathrm{min}$. Uniaxial extension and compression tests as well as cyclic tests were conducted at a displacement rate of 200~mm/min. Three-point bending tests were performed at 3~mm/min with a fixture span of 240~mm. Reconfiguration tests of five-segment multistable structures were conducted using a Thorlabs LTS150C linear translation stage at a speed of 200~mm/min. Micrographs of the multistable conical shell were acquired using a Keyence VHX-7100 digital microscope. The pneumatic pressure source has a maximum gauge pressure of 600~kPa. Negative gauge pressure is generated by a Bacoeng BA1 vacuum pump connected to a vacuum chamber. Positive pressure was controlled by SMC ITV1030 electro-pneumatic regulators, while negative pressure was controlled using a SMC ITV2090 electro-vacuum regulator. The regulators were driven by an Arduino Uno R3 via a Gravity GP8413 I\textsuperscript{2}C DAC module. SMC VDW20GA solenoid valves were used to selectively connect either positive or negative pressure to the output. In total, six independently controlled output channels were available. Forces generated by the pneumatic artificial muscle were measured using a Bota Medusa six-axis force sensor.

% Acknowledgements
\medskip
\textbf{Acknowledgements} \par %delete if not applicable))
The authors gratefully acknowledge the support from the Swiss National Science Foundation (SNSF) through project no.~219336.

% References
\medskip

\bibliographystyle{MSP}
\bibliography{BibTexBibliography}

\end{document}